\documentclass[chicago]{emulateapj}
\usepackage{amsmath}
\usepackage{graphicx}
\usepackage{multirow}
\usepackage[dvips]{color}

\slugcomment{Draft v4 -- 15July2012}
\shortauthors{}

\begin{document}

\title{Milky Way Supermassive Black Hole: Dynamical Feeding from the Circumnuclear Environment}

\author{Hauyu Baobab Liu\altaffilmark{1,2,3}} \author{Pei-Ying Hsieh \altaffilmark{1,4}}  \author{Paul T. P. Ho\altaffilmark{1,2}} \author{Yu-Nung Su \altaffilmark{1}} \author{Melvyn Wright\altaffilmark{5}} \author{Ai-Lei Sun \altaffilmark{6}} \author{Young Chol Minh \altaffilmark{7}}

\affil{$^{1}$Academia Sinica Institute of Astronomy and Astrophysics, P.O. Box 23-141, Taipei, 106 Taiwan}\email{hlu@cfa.havard.edu}

\affil{$^{2}$Harvard-Smithsonian Center for Astrophysics, 60 Garden Street, Cambridge, MA 02138}

\affil{$^{3}$Department of Physics, National Taiwan University, No. 1, Sec. 4, Roosevelt Road, Taipei 106, Taiwan (R.O.C.)}

\affil{$^{4}$Graduate Institute of Astronomy, National Central University, No. 300, \\Jhongda Rd, Jhongli City, Taoyuan County 32001, Taiwan (R.O.C.)}

\affil{$^{5}$Radio Astronomy Laboratory, University of California, Berkeley 601 Campbell Hall, Berkeley, CA 94720, USA}

\affil{$^{6}$Department of Astrophysical Sciences, Peyton Hall, Princeton University, Princeton, NJ, 08544, USA}

\affil{$^{7}$Korea Astronomy and Space Science Institute (KASI), 776 Daeduk-daero, Yuseong, Daejeon 305-348, Korea}

\altaffiltext{1}{Academia Sinica Institute of Astronomy and Astrophysics}
\altaffiltext{2}{Harvard--Smithsonian Center for Astrophysics}
\altaffiltext{3}{Department of Physics, National Taiwan University}
\altaffiltext{4}{Graduate Institute of Astronomy, National Central University}
\altaffiltext{5}{Radio Astronomy Laboratory, University of California, Berkeley}
\altaffiltext{6}{Department of Astrophysical Sciences, Princeton University}
\altaffiltext{7}{Korea Astronomy and Space Science Institute}

\begin{abstract}
The  supermassive black hole (SMBH), Sgr A*, at the Galactic Center is surrounded by a molecular circumnuclear disk (CND) lying between 1.5--4 pc radii. 
The irregular and clumpy structures of the CND, suggest dynamical evolution and episodic feeding of gas towards the central SMBH. 
New sensitive data from the SMA and GBT,  reveal several $>$5--10 pc scale molecular arms, which either directly connect to the CND, or may penetrate inside the CND. 
The CND appears to be the convergence of the innermost parts of largescale gas streamers, which are responding to the central gravitational potential well. 
Rather than being a quasi--stationary structure, the CND may be dynamically evolving, incorporating inflow via streamers, and feeding gas towards the center.
\end{abstract}

\keywords{Galaxy: center --- Galaxy: structure --- Galaxy: kinematics and dynamics --- ISM: clouds}

\clearpage
\section{Introduction }
\label{chap_introduction}
Phenomena associated with supermassive black holes (SMBH), such as  Active Galactic Nuclei (AGN), and circumnuclear starbursts occurring in the innermost few parsecs ($pc$) radii, are intriguing problems in astronomical observations (Antonucci 1993; Davies et al. 2007).
The Milky Way SMBH, Sgr A* (see Genzel et al. 1997 and Ghez et al. 2005 and therein), is the nearest Active Galactic Nucleus (AGN) by a factor of more than 100 (see Morris \& Serabyn 1996 and Mezger et al. 1996 for reviews). 
It is a unique laboratory to spatially resolve dynamical processes within the circumnuclear (parsec scale) environment, which are ultimately responsible for feeding the central engine. 
Surrounding Sgr A*, there are several distinct components: a warm ($>$100 K) molecular circumnuclear disk (CND) residing in the inner 1.5--4 $pc$ radii (G\"{u}sten et al. 1987; Jackson et al. 1993; Marshall et al. 1995; Chan et al. 1997; Christopher et al. 2005; Montero-Casta{\~n}o et al. 2009), several ionized mini--spiral arms converging from the inner edge of the CND toward the Sgr A* (Lo \& Claussen 1983; Roberts \& Goss 1993), and tens of massive OB stars inside the innermost $\sim$0.1 pc region forming rings/disks orbiting about the Sgr A* (Paumard et al. 2006). 
These dynamic components share the common feature of approximate Keplerian motions (G\"{u}sten et al. 1987; Paumard et al. 2006; Zhao et al. 2009, 2010).

A key question is whether all these resolved components in the Galactic Center (GC) are part of a single cohesive dynamical structure. 
Observations of external galaxies detect inflows, which bring molecular gas inward, from $kpc$ scale to the central few tens of $pc$ area (e.g. Davies et al. 2009). 
The resolved GC dynamic components, may define the inflow process from the few tens of $pc$ region exterior of the CND (Sanders 1998; Wardle \& Yusef-Zadeh 2008; Namekata \& Habe 2011). 
The key is to connect the kinematics and the morphology of the molecular CND to the external gas clouds and streamers which follow the Galactic rotational motion. 
Gravitationally captured gas will follow close orbits (i.e. eccentricity $e$$<$1). 
Gas streamers that carry away excess angular momentum, will follow open ($e$$\ge$1) orbits.

Deriving comprehensive images for understanding the structure and dynamics of the Galactic center requires high angular resolution and a careful selection of molecular gas tracers (see also Mart{\'{\i}}n et al. 2012 for discussion about chemistry). 
Previous interferometric observations of the GC already suggested that the high excitation molecular lines in the submillimeter band (e.g. HCN 4--3) are excellent tracers of hot molecular gas around the CND, avoiding confusion from foreground emission and absorption (Jackson et al. 1993; Montero-Casta{\~n}o et al. 2009). 
The abundance of the CS molecule (10$^{-8}$--10$^{-9}$) is known to be only mildly enhanced in UV and shocked environments (Amo-Baladr{\'o}n et al. 2011 and references therein). 
Observations of deeply embedded OB star forming regions in giant molecular clouds suggest that the CS 1--0 transition is a good tracer of the dense ($>$5$\times$10$^{4}$ cm$^{-3}$) molecular filaments feeding the central hot ($>$100 K) molecular cores/toroids (Liu et al. 2011). 
Analogously, we expect the observations of CS 1--0 emission around the CND to reveal the kinematics and morphology of the exterior molecular gas clouds/streamers.

We made wide--field (157 pointings, $\sim$5$'$$\times$5$'$ field of view, $\sim$5$''$ resolution) images of three high--excitation molecular gas tracers ($^{12}$CO 3--2, HCN 4--3, C$^{34}$S 7--6) using the Submillimeter Array (SMA\footnote{The Submillimeter Array is a joint project between the Smithsonian Astrophysical Observatory and the Academia Sinica Institute of Astronomy and Astrophysics, and is funded by the Smithsonian Institution and the Academia Sinica (Ho et al. 2004).}). 
We also made a 20$''$ resolution CS 1--0 image using the National Radio Astronomy Observatory (NRAO\footnote{The National Radio Astronomy Observatory is a facility of the National Science Foundation operated under cooperative agreement by Associated Universities, Inc.}) Robert C. Byrd Green Bank Telescope (GBT). 
The high--excitation lines observed with the SMA, trace the denser and warmer ($>$10$^{5}$ cm$^{-3}$, $>$30 K) gas component, while the CS 1--0 traces the less dense and cooler ($\sim$5$\times$10$^{4}$ cm$^{-3}$, $<$10 K) material.
The new observations are described in Section \ref{chap_obs}.
The results are presented in Section \ref{chap_result}.
A brief discussion is provided in Section \ref{chap_summary}.


\section{Observations and Data Reduction} 
\label{chap_obs}
\subsection{The SMA Observations}
\label{sub_smadata}
We observed high excitation molecular transitions in the submillimeter band using the SMA, in its subcompact array configuration. 
The velocity width of each spectral channel is about 1.4 km\,s$^{-1}$. 
Observations were made in five observing runs, on 2011 June 26, 28 and 29, and July 01 and 03. 
Each run was a mosaic observation of the Galactic Center, with $\sim$4 hours in the target loop. 
The number of available antennas were 6, 7, 6, 7 and 7 in these runs, respectively. 
The minimum and maximum baselines are 5.9 m (7.0 $k\lambda$) and 69.1 m (82 $k\lambda$), respectively. 
Our SMA observations are therefore sensitive to a maximally detectable scale of $\sim$18$''$. 
The system temperature T$_{sys}$ was $\sim$300--600 K,
except for the June 26 run which had poor weather (T$_{sys}$$\sim$600--1200 K).
We observed two phase calibrators 1733--130 and 1924--292 every $\sim$15 minutes in all runs. 
The amplitude and passband calibrators were Titan and 3C 279 in the first run, Neptune and Uranus on June 26, 28, 29, and July 01, and Neptune and 3C 279 on July 03. 
The target loops iterated over 157 pointing centers, with a 5 seconds integration time, and 5 integrations 
at each pointing center. 
Each of the 157 pointings was visited about twice in each observing run (i.e. on source time $\sim$2 hours in total in each run).

We used the MIR IDL software package to edit and calibrate the SMA data. 
About 50\% of the science data on June 29 were flagged because of some unknown abnormal responses in Tsys. 
Data in other runs were not significantly flagged. 
We imaged the HCN 4--3 and C$^{34}$S 7--6 data from the five observing runs together, using the MIRIAD software package (using tasks: \texttt{INVERT}, \texttt{MOSSDI}, \texttt{RESTOR}). 
This yielded a synthesized beamwidth $\theta_{maj}$$\times$$\theta_{min}$  = 5$''$.9$\times$4$''$.4, and BPA. $\sim$18$^{\circ}$; and an RMS noise level $\sim$1100 mJy\,beam$^{-1}$ ($\sim$0.43 K) in each 1.4 km\,s$^{-1}$ line--free channel. 
The $^{12}$CO 3--2 data are more affected by missing short spacing, which can result in significant negative--brightness defects.
We carried out the maximum entropy method (MEM) deconvolution for the dirty image of $^{12}$CO 3--2 using the task \texttt{MOSMEM} in MIRIAD, which helps suppress the defects (see discussion in McGary et al. 2001).
The criterion of the convergence is the ratio of the map noise level and theoretical noise level equal to one. 
The convergent iterations are from 40 to 500, and all the channels converge to the criterion. 
We do not use prior flux model as the initial input.
The $^{12}$CO 3--2 images are only used for qualitative comparison. 

Among the presented submillimeter band transitions in this paper, the $^{12}$CO 3--2 transition is excited by the lowest gas critical density ($\sim$10$^{5}$\,cm$^{-3}$) and the lowest upper--level energy (E$_{up}$$\sim$33 K). The HCN 4--3 and C$^{34}$S 7--6 transitions trace higher temperatures (43 K and 66 K) and higher critical densities ($\sim$10$^{8}$\,cm$^{-3}$ and $\sim$10$^{7}$\,cm$^{-3}$;  Montero-Casta{\~n}o et al. 2009; Evans 1999), which will reveal the embedded dense molecular clumps or the compressed molecular ridges in the $^{12}$CO 3--2 emission gas. 
The C$^{34}$S 7--6 transition traces the highest excitation temperature, and traces the hottest gas around the CND.

 

\begin{figure}[h]
\begin{center}
\includegraphics[width=7cm]{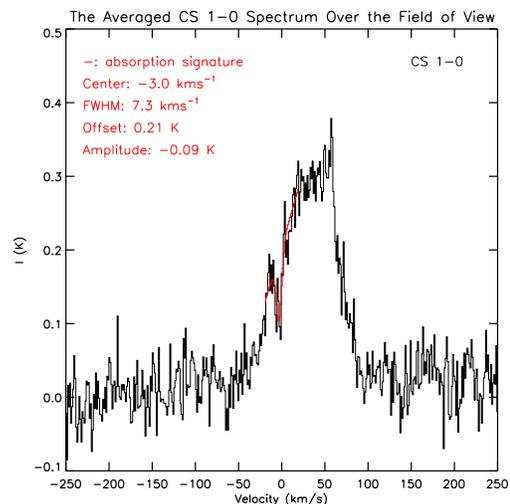}
\end{center}
\vspace{-0.7cm}
\caption{The averaged spectrum of CS 1-0 (smoothed to 1.2\,km\,s$^{-1}$ velocity resolution) over the entire observing field of view. }
\label{fig_allcsspectra}
\end{figure}

\subsection{The GBT Observations}
\label{sub_nh3data}
We observed the CS 1--0 transition using the NRAO GBT. 
The beamwidth is about 15$''$.3. 
Observations were divided into two sessions: on 2011 November 04 with 1.5 hours duration, and on 2011 November 07 with 2.5 hours duration. 
The bright point source 1733--130 was observed in the beginning of each session for antenna surface (i.e. by Out of Focus Holography) and pointing calibrations. 
A line--free reference position RA: 17$^{\mbox{h}}$43$^{\mbox{m}}$43$^{\mbox{s}}$.344, Decl.: -29$^{\circ}$59$'$32$''$.27 was integrated for 30 seconds before and after the target observations in each block for off--source calibration data. 
We used the GBTIDL software package to calibrate the GBT data. 
We used the Astronomical Image Processing System (AIPS) package of NRAO to perform imaging. 
We smoothed the final image to an optimized $\theta_{maj}$$\times$$\theta_{min}$ = 20$''$$\times$18$''$, and BPA. = 0$^{\circ}$ to suppress the striping defects due to the sampling rates. 
The achieved RMS noise level in each 24 kHz (0.15 km\,s$^{-1}$) spectral channel is $\sim$0.53 K (T$_{a}$).

The CS 1--0 transition was selected for our observations because that it is generically optically thin.
The typical Galactic foreground+background $^{12}$CO 1--0 optical depth is 5--50 (c.f. Oka 1998, and references therein).
Our earlier SMA observations of the $^{12}$CO 2--1 emission and the $^{12}$CO 2--1 and $^{13}$CO 2--1 absorption lines against the bright ultracompact H\textsc{ii} region G10.6-0.4 (located in the near side of the Galactic plane) also suggested a foreground $^{12}$CO 2--1 optical depth in the range of 1--10 (the $^{12}$CO 2--1 emission/absorption line is published in Liu, Ho, \& Zhang 2010).
Although the critical density and the upper level energy of CS 1--0 are also low, its Galactic abundance ($\sim$10$^{-9}$; Omodaka et al. 1992) is about 5 orders of magnitude lower than that of $^{12}$CO.
It requires a large amount of gas to be at the same $v_{lsr}$, to develop the optical depth of CS 1--0. 
In the line of sight of the Galactic Center, this $v_{lsr}$ is expected to be $\sim$0\,km\,s$^{-1}$.
Figure \ref{fig_allcsspectra} shows the spectrum of the CS 1--0 line averaged from the entire field of view (see Figure \ref{fig_mnt0}).
The $v_{lsr}$$\sim$0 km\,s$^{-1}$ foreground absorption signature can be seen in Figure \ref{fig_allcsspectra}.
A Gaussian fitting characterizes the FWHM of this absorption signature to be 7.3\,km\,s$^{-1}$, which is narrower than all localized spectral features (e.g. Table \ref{table_gau}). 

Within our observing field of view, previous NRO 45m Telescope observations of CS 1--0 and C$^{34}$S 1--0 derived the optical depth $\tau$(CS 1--0) = 2.8 at the reference point located 3$'$ north and 3$'$ east of the Sgr A* (i.e. in the 50 km\,s$^{-1}$ cloud, see Section 4.1 of Tsuboi et al. 1999). 
Tsuboi et al. (1999) suggested that the CS 1--0 transition is optically thiner in fainter features. 
Strong foreground absorption features are thus not expected (Figure \ref{fig_mnt0}, \ref{fig_csspectra}).



\section{Results}
\label{chap_result}
\begin{figure}[h]
\vspace{-0.5cm}
\hspace{-1cm}
\begin{tabular}{p{3.8cm} p{3.8cm} }
\includegraphics[width=6cm]{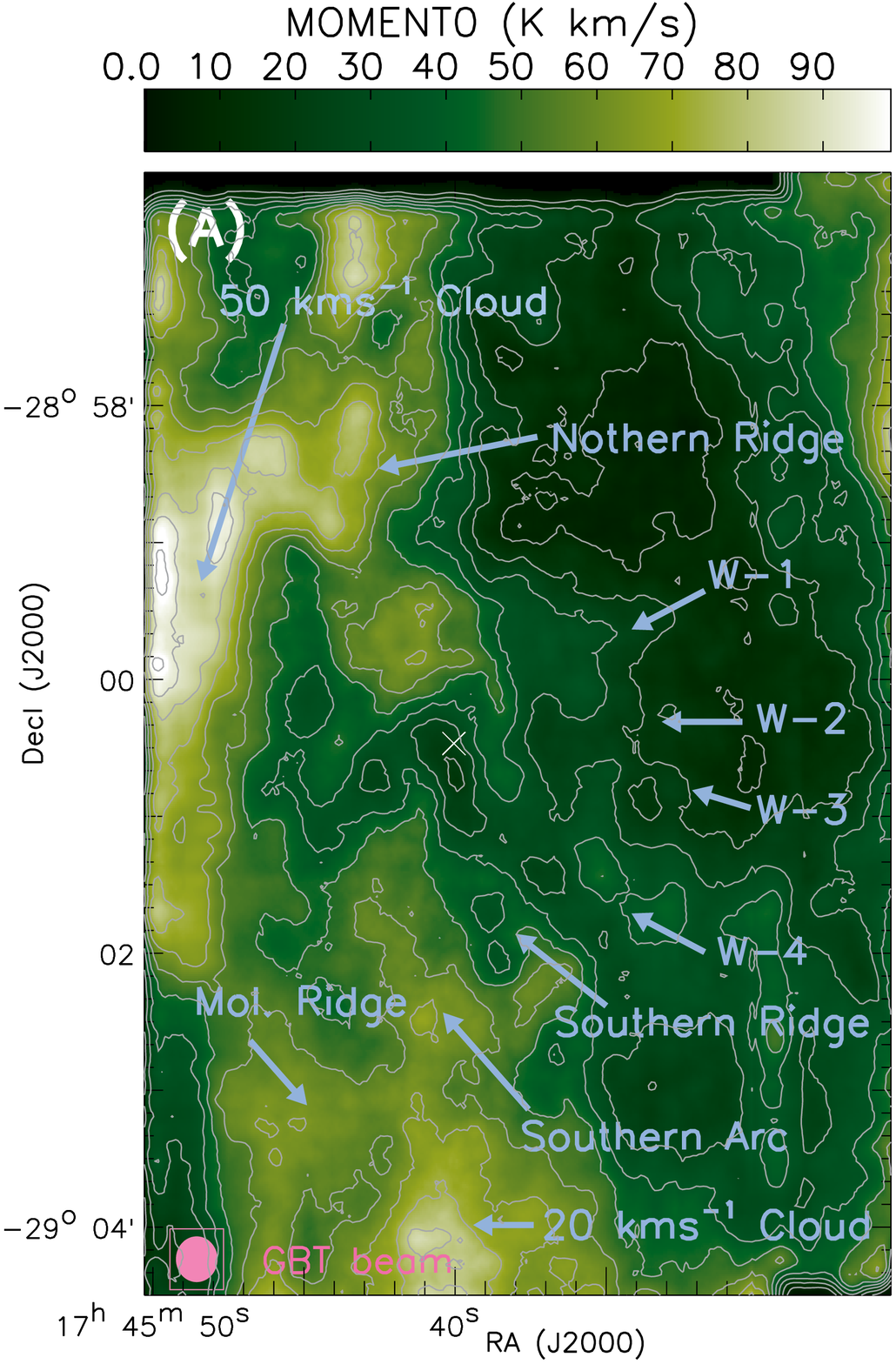} &  \includegraphics[width=6cm]{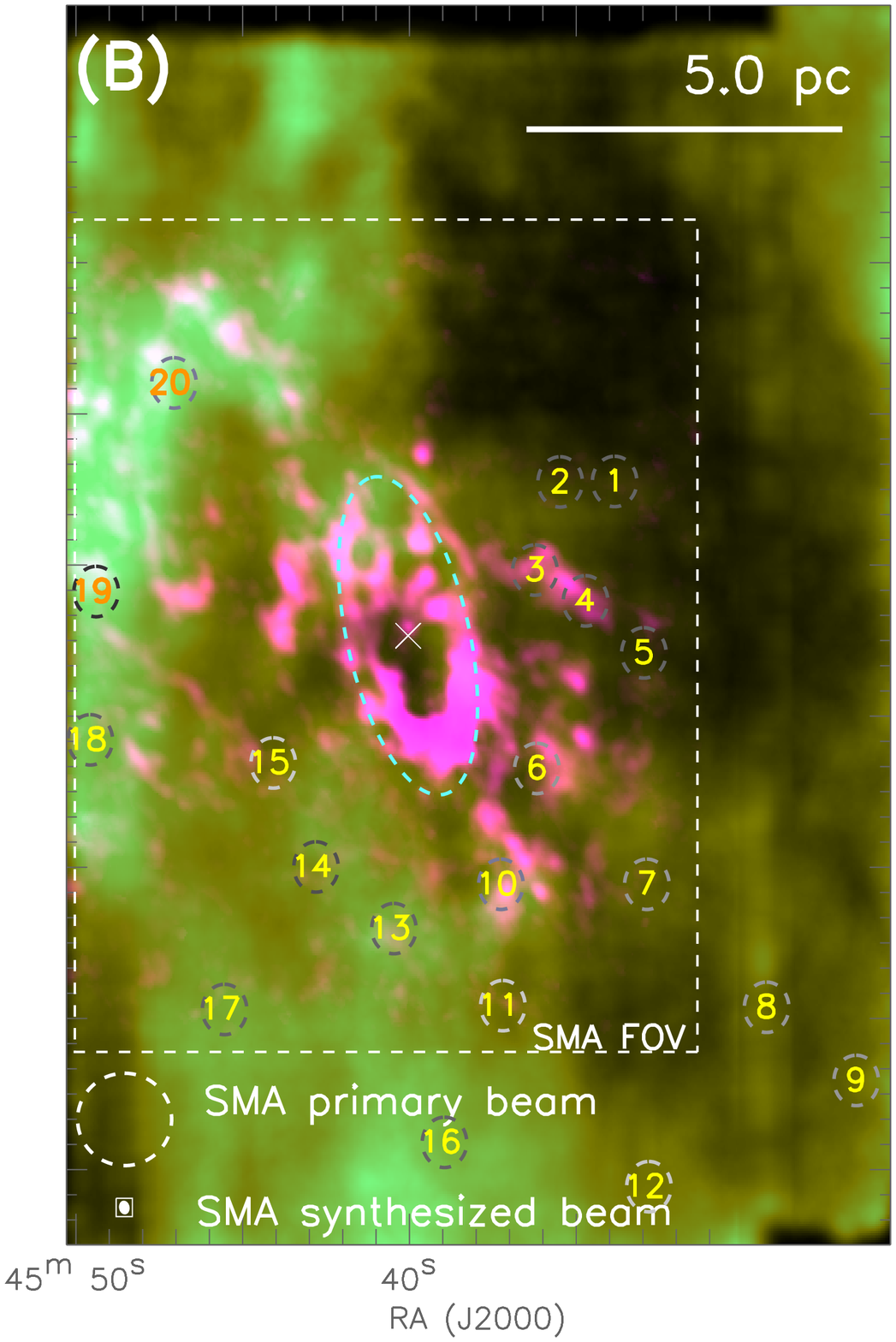}\\
\end{tabular}
\vspace{-0.7cm}
\hspace{-0.5cm}
\caption{\footnotesize{The CND and the exterior molecular streamers in the Galactic Center. 
(A) The velocity integrated CS 1--0 emission. Contour levels are 9.2 K\,km\,s$^{-1}$$\times$[1, 2, 3, 4, 5, 6, 7, 8, 9, 10]. Cross marks the location of Sgr A* (RA: 17$^{\mbox{h}}$45$^{\mbox{m}}$40$^{\mbox{s}}$.04, Decl.: -29$^{\circ}$00$'$28$''$.1). (B) The velocity integrated CS 1--0 emission (green) overlaid with the velocity integrated HCN 4--3 emission (magenta). The box region covers the SMA mapping area (Figure \ref{fig_sma}). A dashed ellipse indicates the CND region. 
The spectra of the marked Region 1--20 are presented in Figure \ref{fig_csspectra}.
The noise scale in the velocity integrated map depends on locally the signal from how many velocity channels is integrated (c.f. Dame 2011).
To give the rough idea of the significance of the integrated CS 1--0 emission, the integration of the signal over a 20 km\,s$^{-1}$ velocity range has rms noise level of 0.92 K\,km\,s$^{-1}$.
}}
\label{fig_mnt0}
\end{figure}

\begin{figure}[h]
\includegraphics[width=14cm]{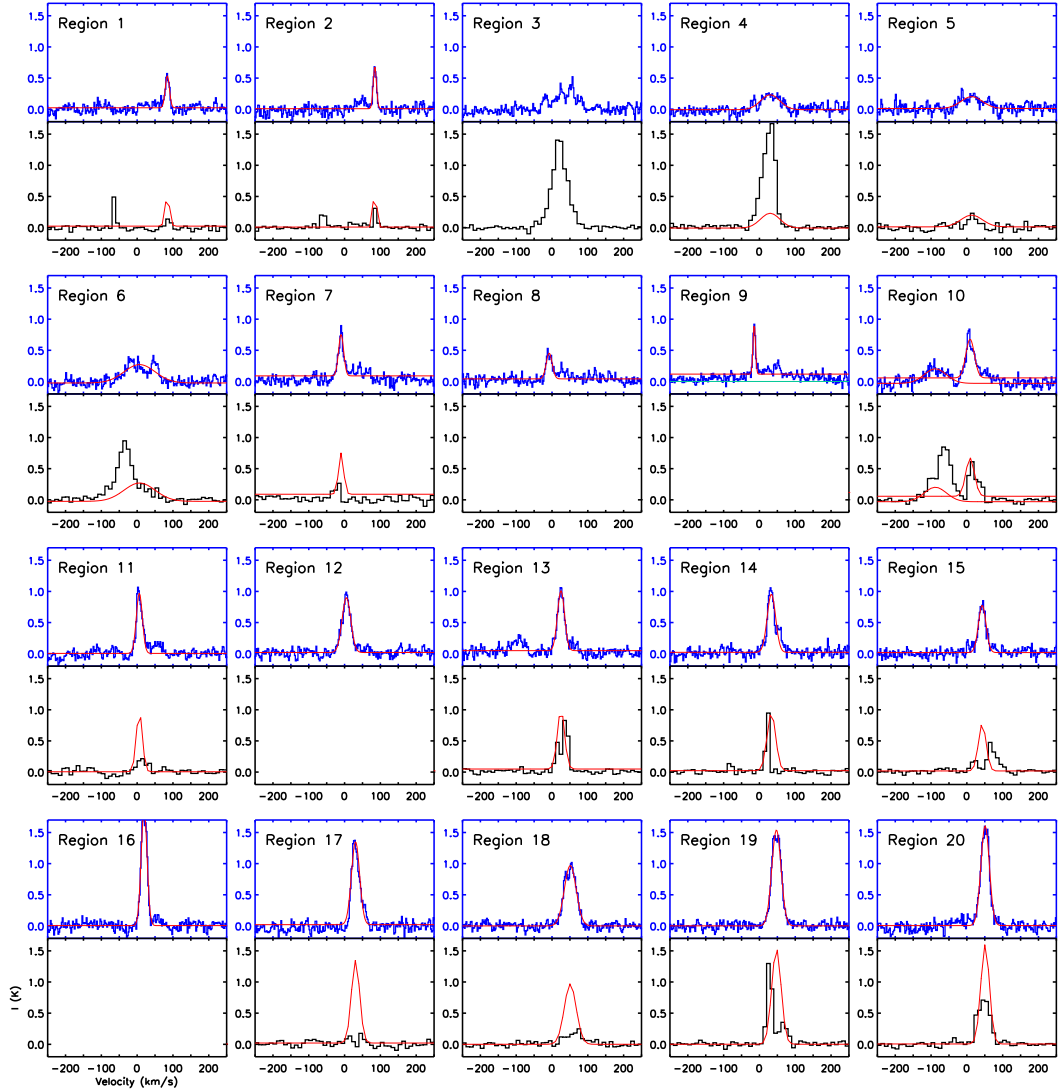}
\caption{The CS 1--0 (blue; smoothed to 2.4\,km\,s$^{-1}$ velocity resolution) and the HCN 4--3 (black) spectra in selected $\theta_{maj}\times\theta_{min}$=20$''$$\times$20$''$ circular regions (see Figure \ref{fig_mnt0}). 
For each region, we characterize the velocity and linewidths of significant CS 1--0 spectral features by Gaussian fitting. 
We present the results of fitting the CS line in both the panels of the CS 1--0 spectrum and the HCN 4--3 spectrum in red.
The Gaussian parameters are summarized in Table \ref{table_gau}. 
We omit fitting the spectrum in Region 3 because of the blending of multiple components. The narrow HCN 4-3 peaks around 65 km\,s$^{-1}$ in Region 1 and 2 appear to be caused by the repetitive stripy defects and should be omitted from the discussion. Region 8, 9 and 16 are located outside of the field of view of our SMA mosaic observations and therefore do not have the measurements of HCN 4--3. The HCN 4--3 spectral profiles in Region 15, 17, 18 and 19 appear to be biased by the missing flux.
}
\label{fig_csspectra}
\end{figure}

\begin{figure}[h]
\rotatebox{-90}{
\includegraphics[width=16cm]{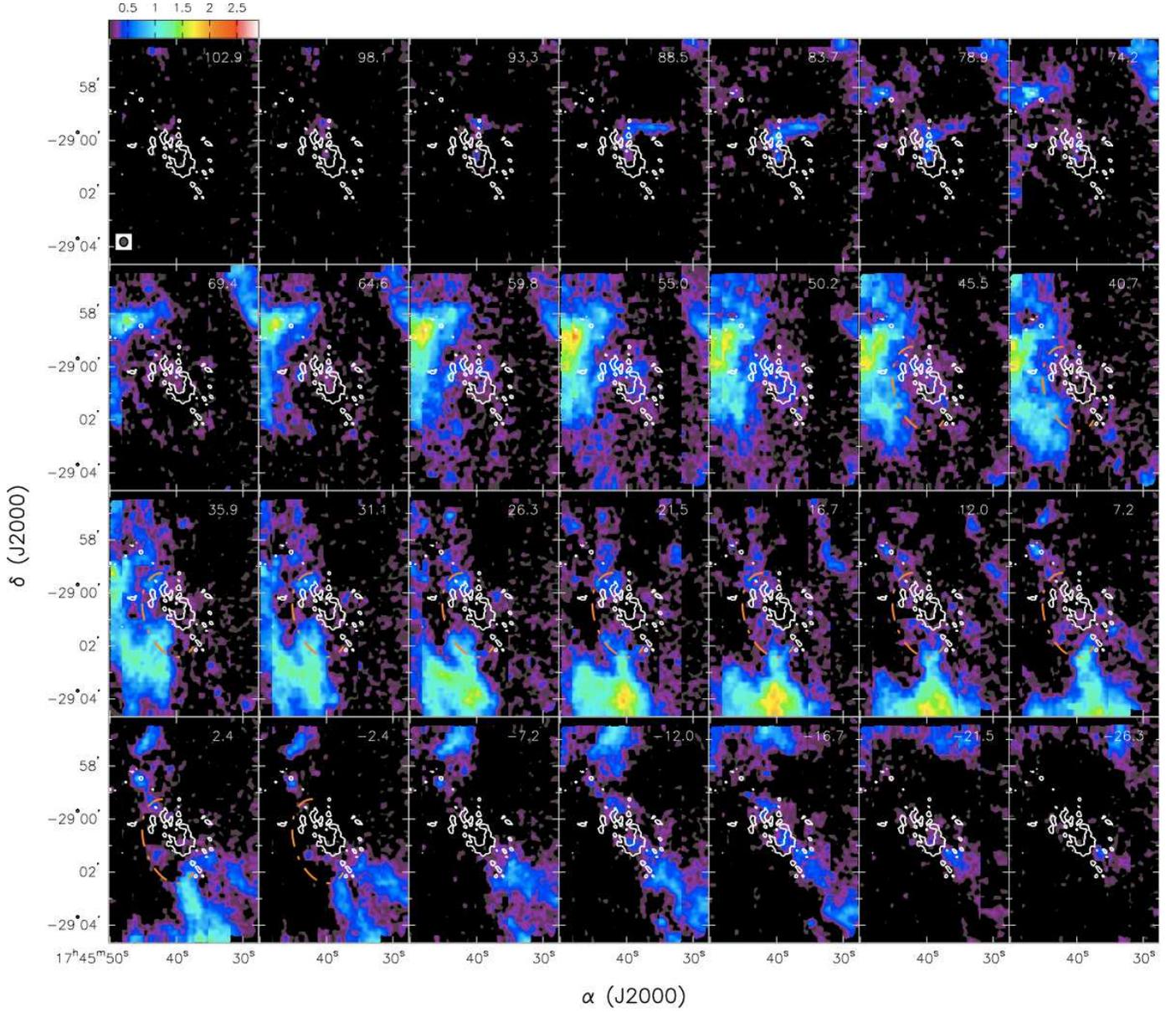} }
\caption{The velocity channel maps of the CS 1--0 line (color; regridded to 4.8\,km\,s$^{-1}$ velocity resolution). The color bar is in Kelvin units. White contours show the 300\,Jy\,beam$^{-1}$\,km\,s$^{-1}$ level of the velocity integrated HCN 4--3 intensity map. The dashed--dotted arcs are drawn to indicate the Southern Arc.}
\label{fig_cschannel}
\end{figure}

\begin{figure}[b]
\hspace{-1cm}
\begin{tabular}{p{3.8cm} p{3.8cm} }
\includegraphics[width=6cm]{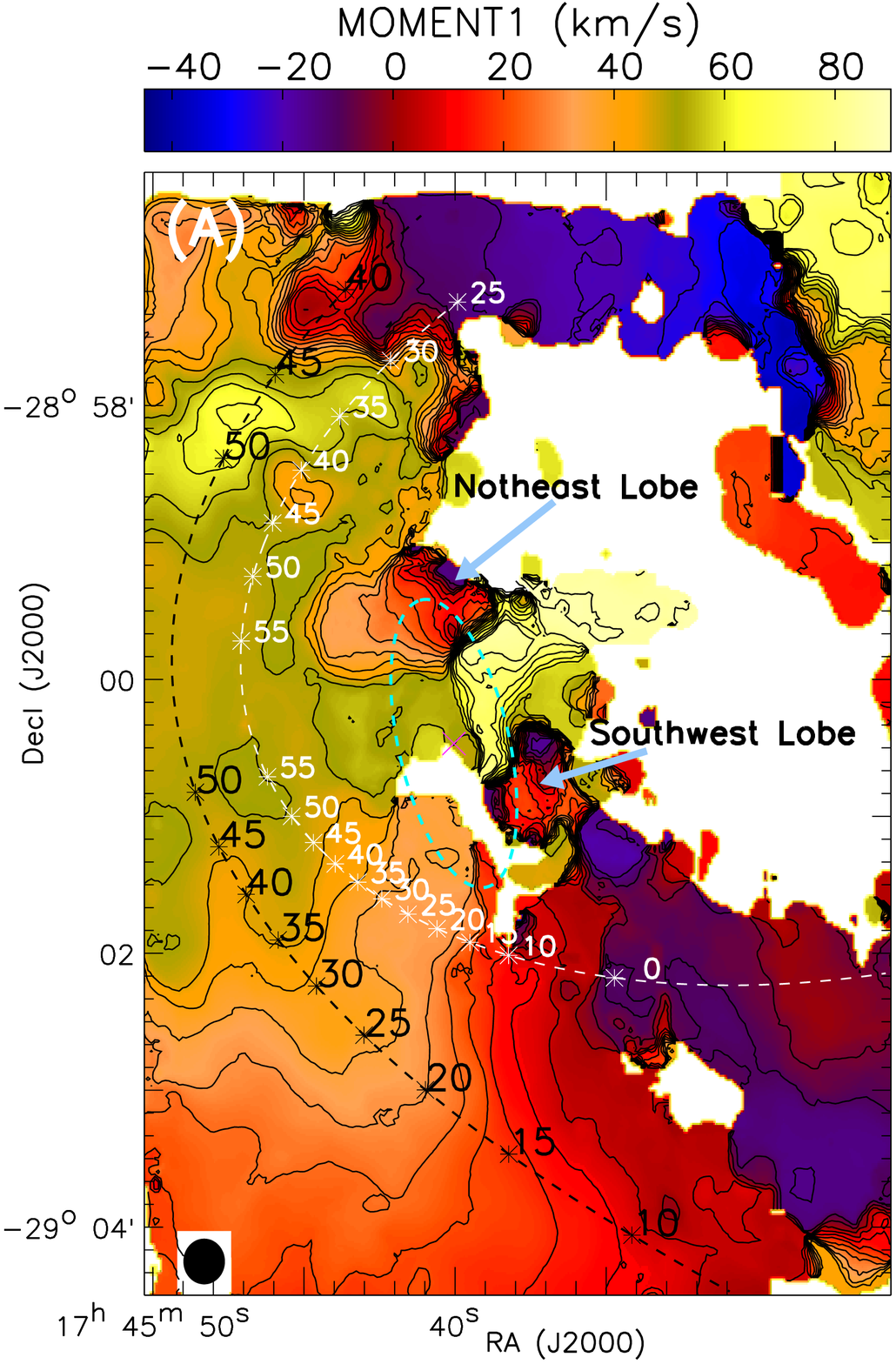} & \includegraphics[width=6cm]{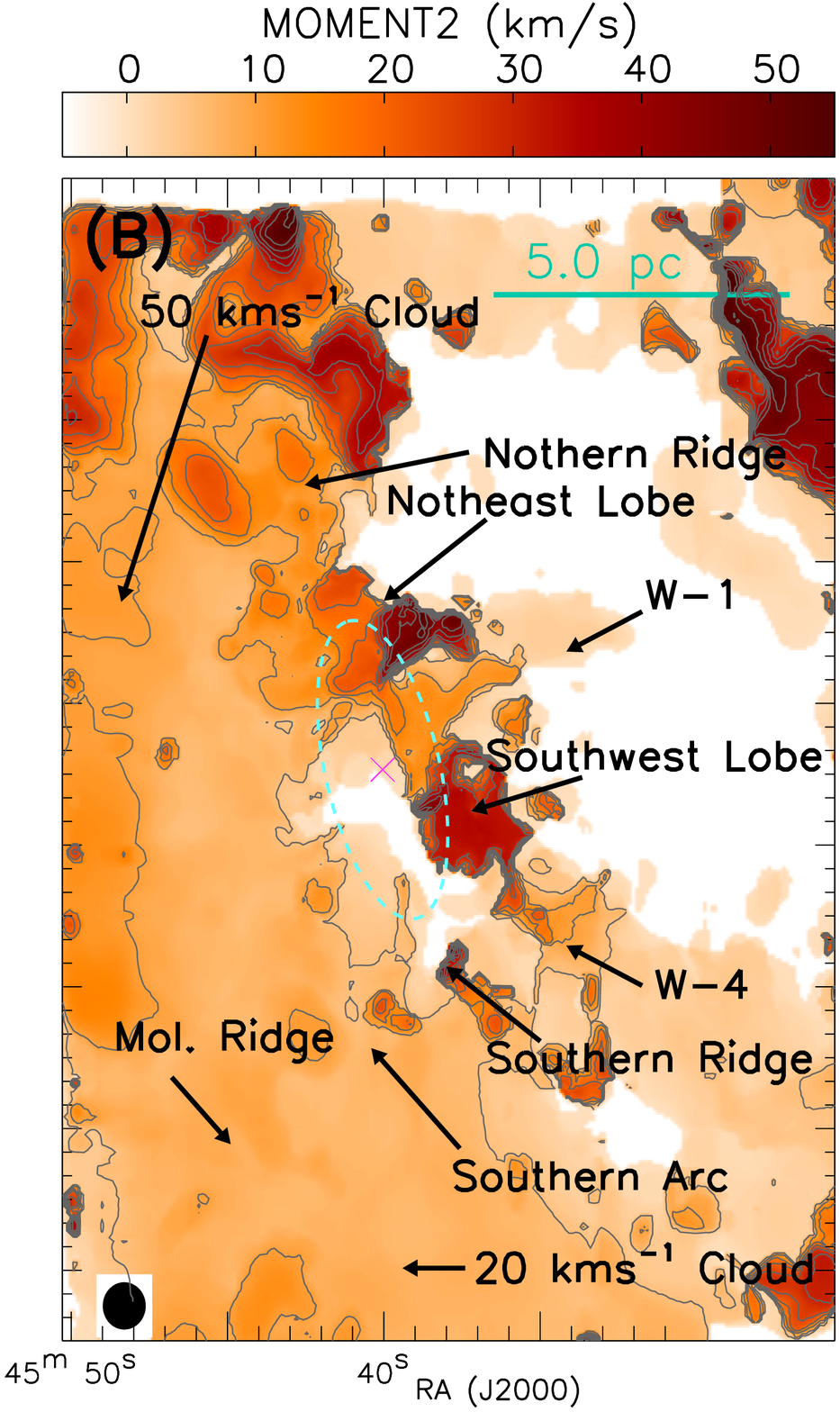}\\
\end{tabular}
\caption{\footnotesize{
The dynamical profile in the Galactic Center. 
(A) The intensity--weighted average velocity map (i.e. moment one map) of CS 1--0. Contour levels start from 0 km\,s$^{-1}$, with intervals of $\pm$5 km\,s$^{-1}$. 
White and black dashed lines show two examples of Keplerian orbits: one pass $\textbf{x}$ = (-3.3, 3.5, - 2.2) $pc$, $\textbf{v}$ = (-24,-62,47) km\,s$^{-1}$; the other pass $\textbf{x}$ = (-4.0, 5.0, - 2.7) $pc$, $\textbf{v}$ = (-28,-55,50) km\,s$^{-1}$. 
Positive xyz directions are defined in west, north and in the line--of--sight, relatively to the location of the Sgr A*. 
The line--of--sight velocities (km\,s$^{-1}$) of these orbits are labeled with star symbols.  
(B) The intensity--weighted velocity dispersion map (i.e. moment two map) of CS 1--0. Contour levels start from 5 km\,s$^{-1}$, with intervals of 5 km\,s$^{-1}$. 
The angular resolution of the CS 1--0 maps is shown in the bottom left of both panels. 
The dashed ellipses indicate the CND region.
}}
\label{fig_mnt12}
\end{figure}

Our GBT and SMA images recover all the structures detected in previous observations (Figure \ref{fig_mnt0}; nomenclature follows Christopher et al. 2005 and Amo-Baladr{\'o}n et al. 2011). 
In addition, our high angular resolutions reveal more detailed kinematics and morphology, while our high sensitivity allows the denser structures to be seen in the context of fainter features. 
Structures previously known as the 50 km\,s$^{-1}$ Cloud (c.f. G\"{u}sten \& Downes 1980), the Molecular Ridge, and the 20 km\,s$^{-1}$ Cloud, are seen to be an integrated $\sim$20 $pc$ scale molecular arm winding around the CND. 
Between this $\sim$20 $pc$ scale molecular arm and the CND (see also Figure \ref{fig_cschannel}), we resolve another arc--shaped molecular streamer which is most manifest toward the south of the CND (Southern Arc hereafter). 
Both the $\sim$20 $pc$ scale molecular arm and the Southern Arc follow smooth velocity gradients, continuously from 50 km\,s$^{-1}$ on the northeast side of the CND, to $\le$20 km\,s$^{-1}$ on the south side of the CND, to $\sim$10 km\,s$^{-1}$ towards the southwest (Figure \ref{fig_csspectra}, \ref{fig_mnt12}, Table \ref{table_gau}). 
The velocity dispersion in these structures is around 5--10 km\,s$^{-1}$ and is relatively uniform (Figure \ref{fig_mnt12}B).
Given the critical volume density $n^{\mbox{\scriptsize{CS}}}_{c}\sim5\times10^{4}$\,cm$^{-3}$ to excite CS 1--0 (Evans 1999), and assuming the local structures are approximately spherical, we estimate the 3$\sigma$ significance of the velocity (dispersion) change (3$\Delta v$) based on the local gravitational perturbation:
\begin{equation}
\frac{G\cdot\frac{4}{3}\cdot\pi\cdot(\theta^{\mbox{\scriptsize{CS}}}/2)^3 \cdot n^{\mbox{\scriptsize{CS}}}_{c}\cdot \mu\cdot m_{\mbox{\scriptsize{H} } }}{ \theta^{\mbox{\scriptsize{CS}}}/2 } \sim \frac{1}{2}(\Delta v)^{2},
\end{equation}
where $G$ is the gravitational constant, $\theta^{\mbox{\scriptsize{CS}}}$ is the physical length scale of the beam of our CS 1--0 observations, $\mu$ is the mean molecular weight, and $m_{\mbox{\scriptsize{H}}}$ is the H atom mass.
We suggest that the variations of the velocity larger than 3$\Delta v$$\sim$5.1 km\,s$^{-1}$ value should be attributed to the large--scale kinematics. 
Local increases in velocity dispersion larger than $\sim$5.1 km\,s$^{-1}$ could be a signature of the blending of distinct gas components (e.g. near the Northern and the Southern Ridges), or because of  localized interactions (e.g. shock, supernovae, etc).
The Southern Arc, also noted as a northern extension from the 20 km\,s$^{-1}$ Cloud in early studies (Ho et al. 1991; Okumura et al. 1991), and the $\sim$20 $pc$ scale molecular arm, are most likely part of the same structure.

\clearpage
\clearpage

\clearpage

\clearpage

The resolved overall velocity field is consistent with gravitationally accelerated rotational motions around the central SMBH. 
Here we present two examples of Keplerian orbits generated from $\textbf{a}$ = $-\textbf{x}(GM/|\textbf{x}|^{3})$, $\dot{\textbf{x}}$ = $\textbf{v}$, and $\dot{\textbf{v}}$ = $\textbf{a}$, in Figure \ref{fig_mnt12}: one pass $\textbf{x}$ = (-3.3, 3.5, - 2.2) $pc$, $\textbf{v}$ = (-24,-62,47) km\,s$^{-1}$; the other pass \textbf{x} = (-4.0, 5.0, - 2.7) $pc$, $\textbf{v}$ = (-28,-55,50) km\,s$^{-1}$. 
When generating these Keplerian orbits, we used a simplified assumption of a single point--like gravitational potential source, locating at \textbf{x} = (0.0, 0.0, 0.0) $pc$. 
The mass of the gravitational potential source is 5$\times$10$^{6}$ M$_{\odot}$. 
When test particles pass the coordinates specified above, their kinetic energy will be 79.6\% and 99.5\% of the gravitational potential energy $(GM/r)$, respectively. 
Though these ratios are higher than the ratio of 50\% for circular orbits (i.e. $GM/r^{2} = v^{2}/r$), these Keplerian orbits are still bound by the central gravity. 
We think these orbits are representative of the observed velocity gradients in CS 1--0 emission. 
For example, they reproduce the $\sim$5 km\,s$^{-1}$\,arcmin$^{-1}$ velocity gradient in the east, and the $\gtrsim$15 km\,s$^{-1}$\,arcmin$^{-1}$ velocity gradient in the southeast (Figure \ref{fig_mnt12}). 
Improvements in modeling will require more accurate models of mass distribution, and consideration of hydrodynamical and selfÐ gravitational interactions. 
Given the typical cloud mass of $\sim$10$^{4}$ M$_{\odot}$ at the few parsec scale, we expect localized supersonic motions of plus--minus several km\,s$^{-1}$ to confuse the global orbital motions. 
The impulse from the supernovae will also disturb the motions.
Observationally, we note that the low excitation transition CS 1--0  traces the large--scale motions of the bulk of gas. 
Because of the higher excitation conditions and the interferometric filtering, the molecular line images obtained from the SMA are more sensitive to local kinematics, motions influenced by supernovae, or shocks. 
Detailed discussion of these effects is beyond the scope of the present paper, and will be addressed in future work. 

Significant deviations (i.e., $>$5 km\,s$^{-1}$) from the smooth velocity gradients are resolved (Figure \ref{fig_mnt12}).
Perpendicular to the 15--35 km\,s$^{-1}$ sector of the black dashed orbit, we see outward velocity gradients of -5--\,-10\,km\,$^{-1}$, and  inward velocity gradients of -10--\,-15\,km\,s$^{-1}$, constituting "wiggles" in the iso--velocity contours. 
Geometrically, these "wiggles" resemble those commonly seen for motions across the spiral arms in external galaxies (e.g. Aalto et al. 1999). 
In addition, we see an east--west orientated, blue--shifted ($\sim$ -10 km\,s$^{-1}$-- -30 km\,s$^{-1}$) gas streamer in the north, the compressed velocity contours around the CND, and a complex velocity pattern northeast of the CND. 

Our new results show that the CND is well connected to the external environment (see also Coil \& Ho 1999; Coil \& Ho 2000; McGary et al. 2001; Herrnstein \& Ho 2002; Herrnstein \& Ho 2005; Donovan et al. 2006; Oka et al. 2011; Ferriere 2012). 
The CND is not prominent in the CS 1--0 map (Figure \ref{fig_mnt0}A) because of the higher excitation in the CND. 
The filamentary features which connect the CND to the outside are clearly seen in the SMA maps, and more faintly in the CS 1--0 map (Figure \ref{fig_mnt0}, \ref{fig_cschannel}). 
West of the CND, we resolve the previously known Western Streamers into 4 molecular arms, labeled by W--1,$_{\cdots}$,4. 
The Southern Ridge, the Southern Arm, the Northern Ridge, and the Northeast Lobe, are all very similar to the W--1,$_{\cdots}$,4 arms, in apparently ending on the CND.

\begin{table}[h]\scriptsize{
\hspace{-0.3cm}
\begin{tabular}{ ccc | rccc  } \hline
 \multicolumn{3}{c|}{Region}	       	&	\multicolumn{4}{c}{CS 1--0}		   \\\hline
No.  	& RA & Decl.	& $B$ & $A$	&	$v_{0}$	& FWHM  	\\
	&	(J2000)	&	(J2000)		&	(mK)	 & (K) 	&	(km\,s$^{-1}$)	& (km\,s$^{-1}$)   \\\hline
1	& 17$^{h}$45$^{m}$33$^{s}$.85 &	-28$^{\circ}$59$'$26$''$.8	&  23		&	0.52	&	84.3	&	14\\
2	&	17$^{h}$45$^{m}$35$^{s}$.48	&	-28$^{\circ}$59$'$23$''$.0	&	8.1	&	0.67	&	84.5	&	11\\
3	&	17$^{h}$45$^{m}$36$^{s}$.24	&	-29$^{\circ}$00$'$02$''$.0	&	--- 		&	---		&	---		&	---\\
4	&	17$^{h}$45$^{m}$34$^{s}$.72	&	-29$^{\circ}$00$'$14$''$.2	&	-7.5	&	0.24	&	30.0	&	63\\
5	&	17$^{h}$45$^{m}$32$^{s}$.97	&	-29$^{\circ}$00$'$34$''$.9	&	12		&	0.19	&	13.3	&	71\\
6	&	17$^{h}$45$^{m}$36$^{s}$.18	&	-29$^{\circ}$01$'$20$''$.8	&	-2.4	&	0.29	&	6.8	&	102\\
7	&	17$^{h}$45$^{m}$32$^{s}$.88	&	-29$^{\circ}$02$'$06$''$.7	&	90		&	0.67	&	-9.7	&	17\\
8	&	17$^{h}$45$^{m}$29$^{s}$.29	&	-29$^{\circ}$02$'$55$''$.6	&	42		&	0.41	&	-7.9	&	20\\
9	&	17$^{h}$45$^{m}$26$^{s}$.60	&	-29$^{\circ}$03$'$24$''$.6	&	120	&	0.80	&	-14.5	&	8\\
10	&	17$^{h}$45$^{m}$37$^{s}$.26	&	-29$^{\circ}$02$'$06$''$.7	&	58		&	0.62	&	9.0	&	28\\
	&						&							&  -30	&	0.23	&	-88.1	&	73\\
11	&	17$^{h}$45$^{m}$37$^{s}$.20	&	-29$^{\circ}$02$'$54$''$.9	&	5.1	&	0.97	&	6.1	&	20\\
12	&	17$^{h}$45$^{m}$32$^{s}$.82	&	-29$^{\circ}$04$'$06$''$.8	&	19		&	0.81	&	5.1	&	28\\
13	&	17$^{h}$45$^{m}$40$^{s}$.47	&	-29$^{\circ}$02$'$24$''$.3	&	49		&	0.96	&	25.1	&	23\\
14	&	17$^{h}$45$^{m}$42$^{s}$.80	&	-29$^{\circ}$01$'$59$''$.9	&	22		&	0.93	&	33.3	&	28\\
15	&	17$^{h}$45$^{m}$44$^{s}$.08	&	-29$^{\circ}$01$'$18$''$.6	&	18		&	0.76	&	42.9	&	25\\
16	&	17$^{h}$45$^{m}$38$^{s}$.95	&	-29$^{\circ}$03$'$49$''$.2		&	5.4	&	1.9	&	19.4	&	19\\
17	&	17$^{h}$45$^{m}$45$^{s}$.54	&	-29$^{\circ}$02$'$56$''$.4	&	20		&	1.3	&	30.6	&	27\\
18	&	17$^{h}$45$^{m}$49$^{s}$.57	&	-29$^{\circ}$01$'$09$''$.4	&	0.34	&	0.97	&	51.0	&	39\\
19	& 17$^{h}$45$^{m}$49$^{s}$.39	&	-29$^{\circ}$00$'$10$''$.5	&	1.4	&	1.5	&	47.4	&	31\\
20	&	17$^{h}$45$^{m}$47$^{s}$.06	&	-28$^{\circ}$58$'$48$''$.7	&	5.3	&	1.6	&	51.0	&	28\\\hline
\end{tabular} }
\caption{\footnotesize{The parameters of the Gaussian fittings for spectra in Region 1--20 (Figure \ref{fig_mnt0}, \ref{fig_csspectra}). The Gaussian profile has the form of ($A\,e^{-\frac{(v-v_{0})^{2}}{2\sigma^{2}}}+B$). The full width at half maximum (FWHM) of the Gaussians was obtained by FWHM$\sim$2.35$\times$$\sigma$.}
}
\label{table_gau}
\end{table}

\begin{figure}
\vspace{-0.2cm}
\hspace{-0.3cm}
\begin{tabular}{ p{3.6cm} p{3.6cm} }
\includegraphics[width=5.4cm]{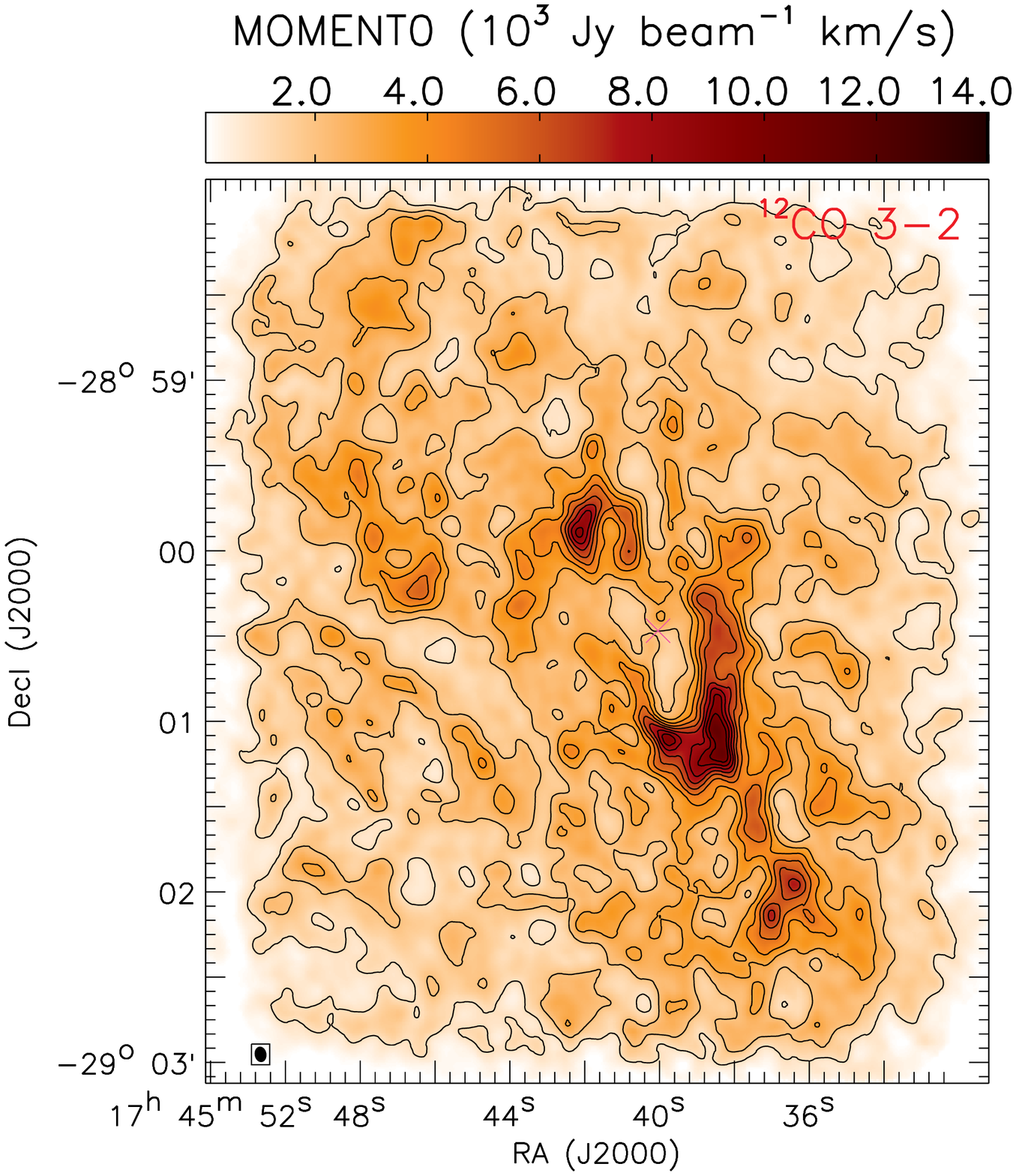} & \includegraphics[width=5.4cm]{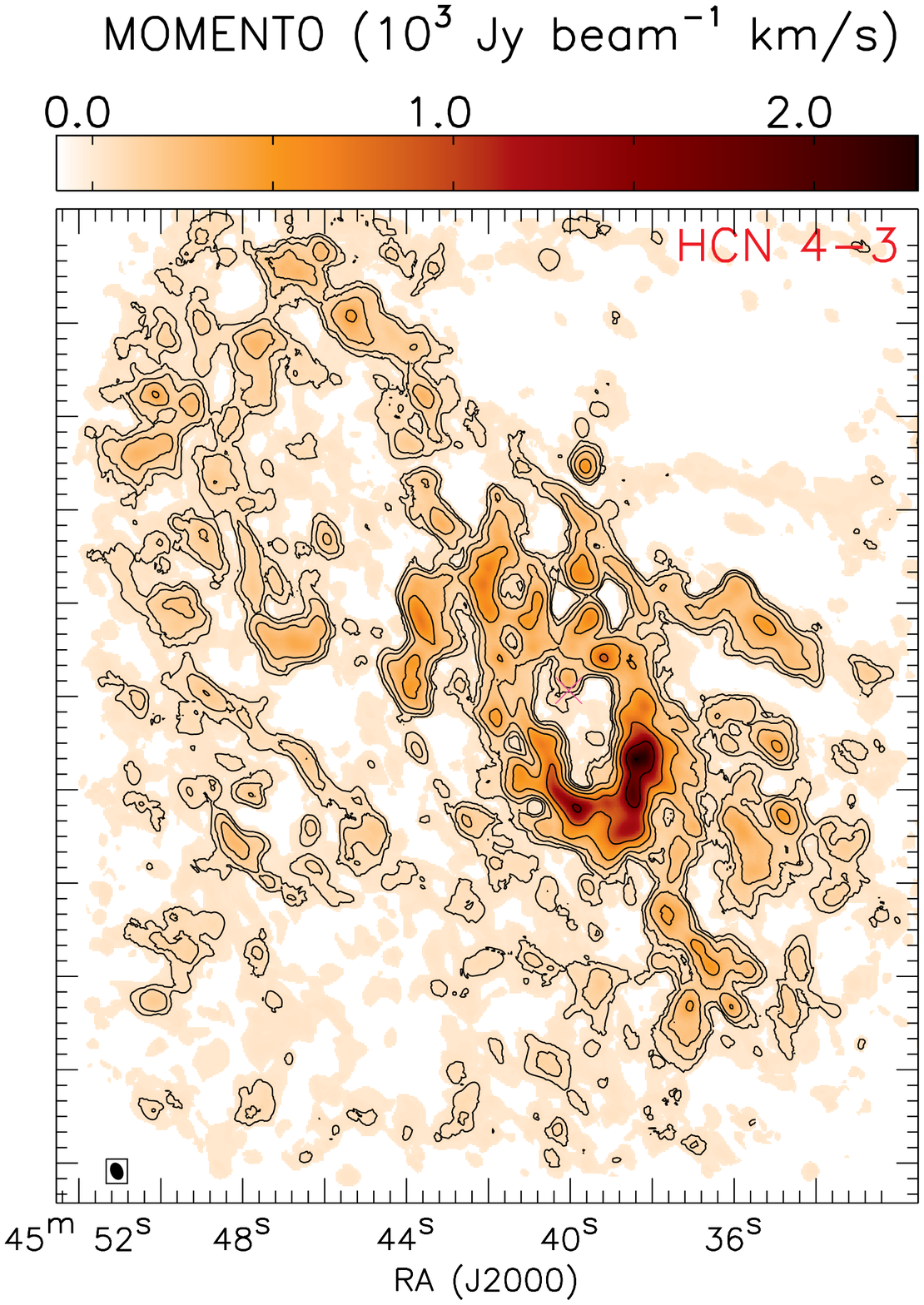}  \\
\end{tabular}

\vspace{-0.8cm}

\hspace{-0.3cm}
\begin{tabular}{ p{3.6cm} p{3.6cm} }
\includegraphics[width=5.4cm]{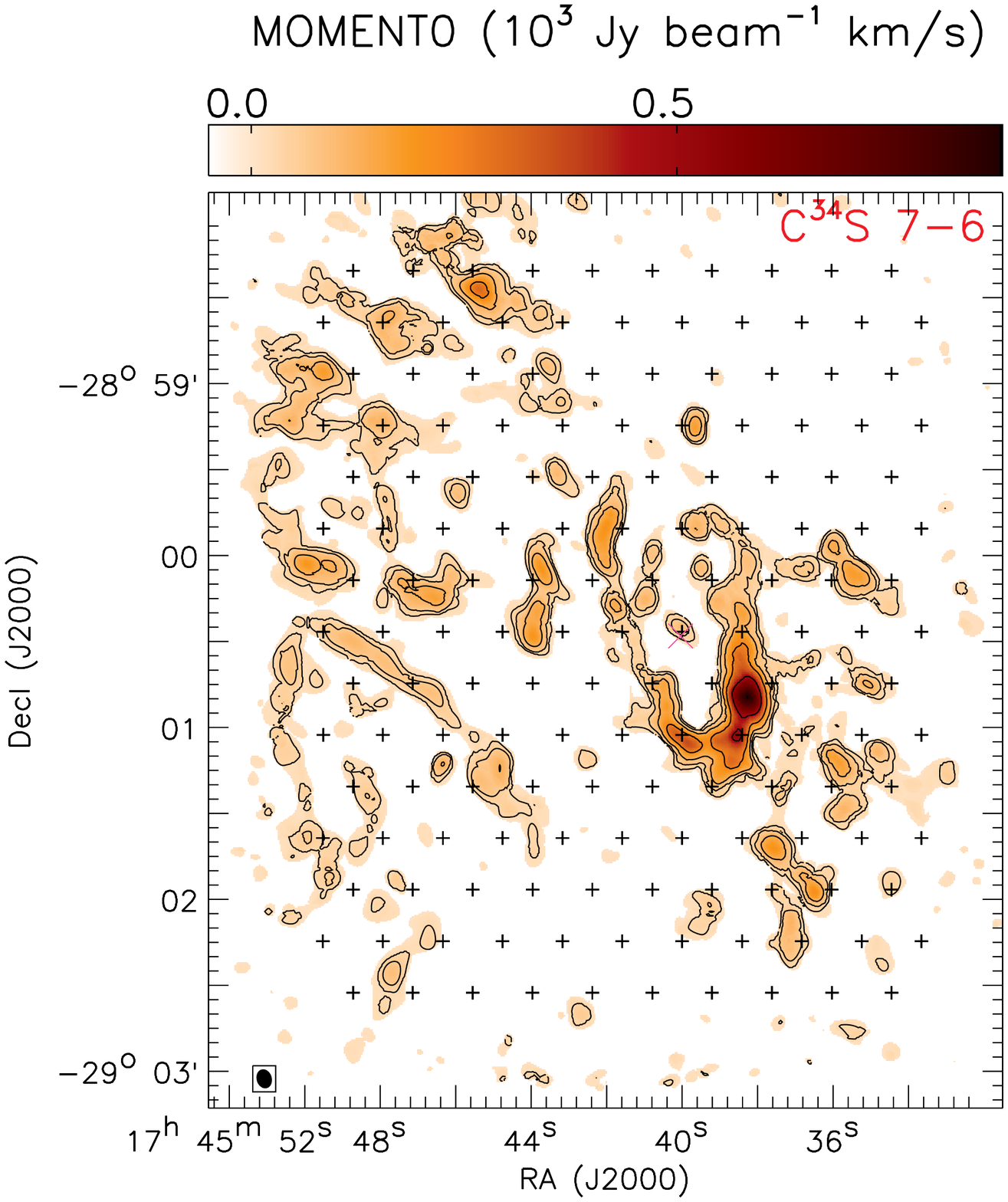} & \includegraphics[width=5.4cm]{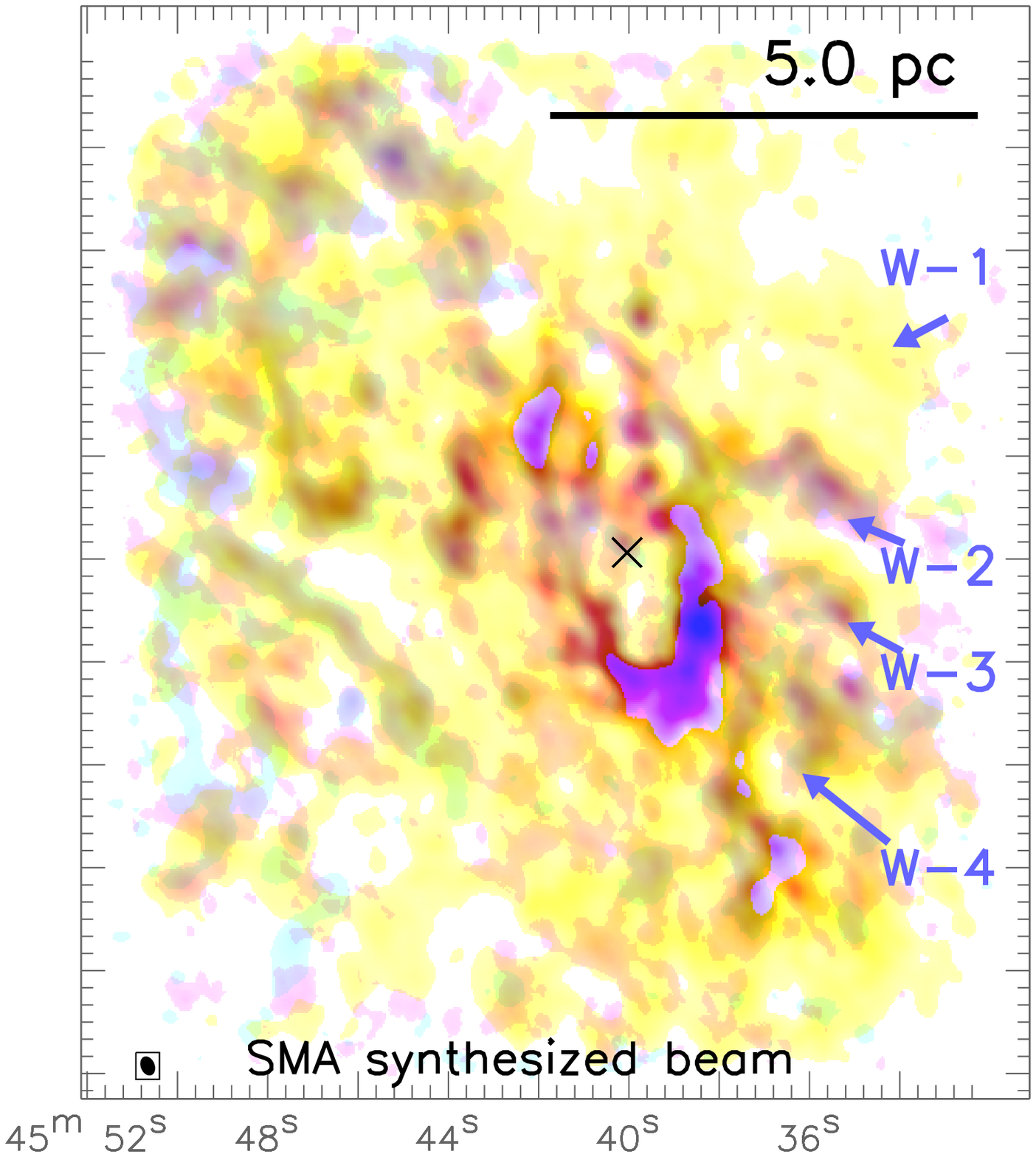}  \\
\end{tabular}
\caption{\footnotesize{Velocity integrated (i.e., moment 0) images of the $^{12}$CO 3--2, HCN 4--3, and C$^{34}$S 7--6 transitions.
The synthesized beam of the SMA observations is shown in the bottom left. 
The contours of the $^{12}$CO 3--2 image start at the value 1000\,Jy\,beam$^{-1}$km\,s$^{-1}$, and are drawn at intervals of 1000\,Jy\,beam$^{-1}$km\,s$^{-1}$.
The contours of the HCN 4--3 and C$^{34}$S 7--6 images are 50\,Jy\,beam$^{-1}$km\,s$^{-1}$$\times$[1, 2, 4, 8, 16, 32] and 30\,Jy\,beam$^{-1}$km\,s$^{-1}$$\times$[1, 2, 4, 8, 16], respectively.
Integration of the signal over a 20 km\,s$^{-1}$ velocity range has an rms noise level of 5.8 Jy\,beam$^{-1}$km\,s$^{-1}$ (2.2 K\,km\,s$^{-1}$).
The bottom right panel shows an overlay of these lines, in yellow ($^{12}$CO 3--2), magenta (HCN 4--3) and cyan (C$^{34}$S 7--6) colors.
Crosses in the C$^{34}$S image mark the pointing centers of the SMA mosaic observations.}}
\label{fig_sma}
\end{figure}


\begin{figure}
\includegraphics[scale=0.5]{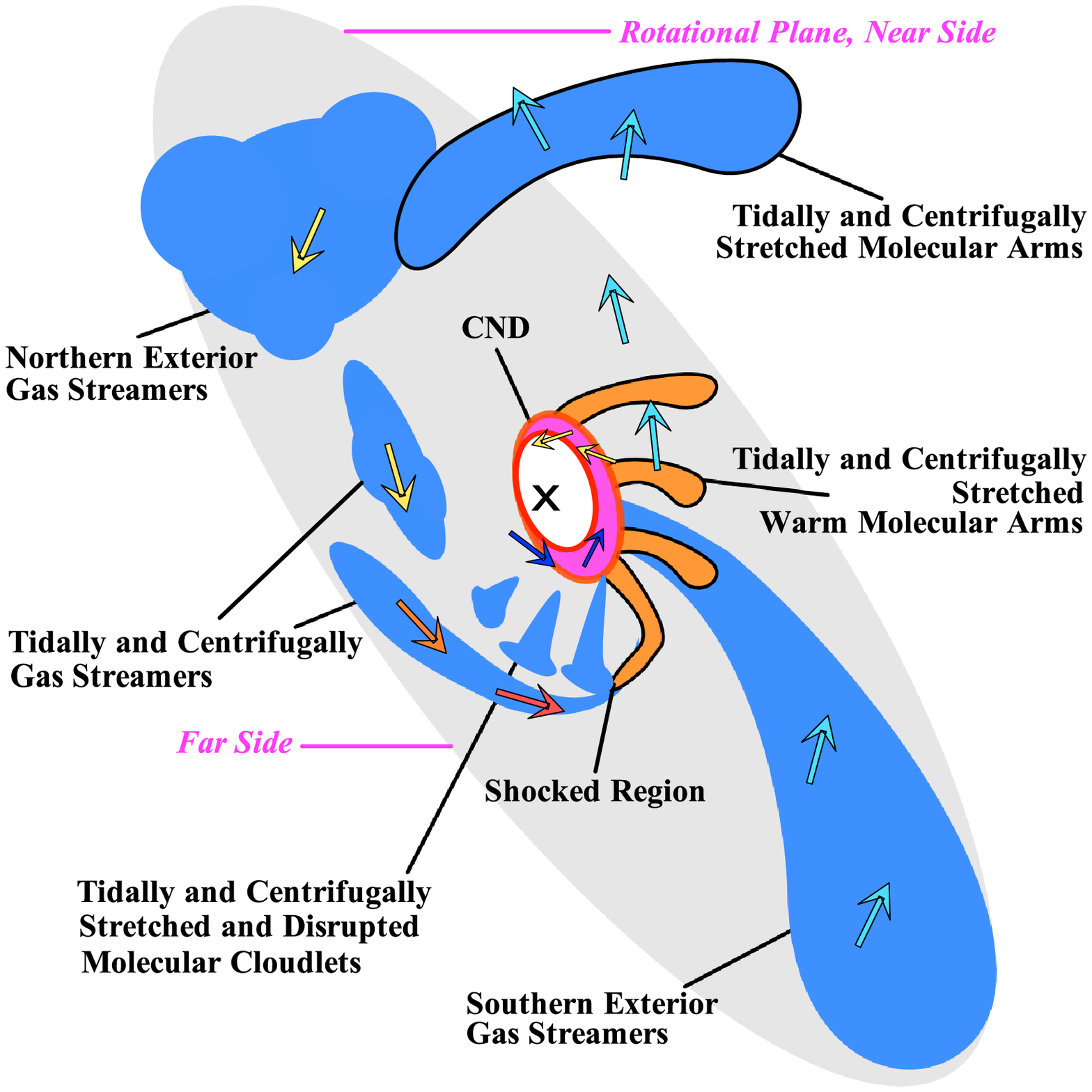} 
\caption{\footnotesize{A schematic picture of the proposed kinematic processes.  Shapes without heavily outlined boundaries, represent gas structures approaching the central black hole; shapes bounded by black lines represent gas structures receding from the central black hole. The magenta shape bounded by red lines represents the hot CND around the central black hole. Arrows indicate the velocity field.  Structures moving toward the near side are blue--shifted. 
The blue and the orange color of shapes represent cooler and warmer gas temperature, respectively.
In this model, dynamical interactions of congregated dense gas around the center redistribute the kinetic energy and angular momentum, and also convert some kinetic energy into heat. 
After interactions, the gas which carries an excess of kinetic energy and angular momentum will be further stretched by the centrifugal force, and then eventually leave the system following open (i.e. eccentricity e$\ge$1) orbits. 
The gas which lost significant kinetic energy and angular momentum during the interactions will be bound by the gravitational potential energy of the central black hole, and may form temporarily a warm CND while continuing inward (c.f. Namekata \& Habe 2011).}}
\label{fig_schematic}
\end{figure}

Furthermore, Figure \ref{fig_mnt12}B shows that the velocity dispersion is enhanced to $\gg$15 km\,s$^{-1}$ at the points where the filaments meet the CND. 
Close to the CND, the high excitation SMA lines (e.g., HCN 4--3), show that protrusions in the CND are connected to external filaments (Figure \ref{fig_sma}). 
The W--1 arm appears to intersect the previously known Northern Ridge at the Northeast Lobe. 
The W--3 and W--4 arms appear to intersect the CND at the Southwest Lobe (see also Dent et al. 1993, Sato \& Tsuboi 2008). 
The Southern Ridge also ends on the southern part of the CND. 
The W--1 arm has an redshifted velocity of $\sim$84.4\,km\,s$^{-1}$ (Figure \ref{fig_mnt0}, Table \ref{table_gau}), and converges to a velocity of 102.9\,km\,s$^{-1}$ at its conjunction with the northern part of the CND (Figure \ref{fig_cschannel}).
Near the conjunction of the Southern Ridge with the CND (e.g. Region 10), we see a broad blueshifted component at -88.1\,km\,s$^{-1}$ (Figure \ref{fig_csspectra}, Table \ref{table_gau}), which is consistent with the blueshifted rotational motion of the CND (Christopher et al. 2005). 
A narrower 9\,km\,$^{-1}$ spectral component is blended in the same region, which may be contributed by the Southern Arc (Figure \ref{fig_mnt12}).
Region 6, which is located near the conjunction of the W--4 arm with the CND, shows a similar spectral profile to Region 10, although it is more difficult to separate the blueshifted and the redshifted components. 
Away from the CND in the W--4 arm (e.g. following Region 7, 8, 9), the blueshifted spectral component becomes at least 2 times narrower and is distinct from the blended
broad line (Table \ref{table_gau}). 
In Figure \ref{fig_csspectra}, we observe broad linewidth components in the western streamers (e.g. Regions 2--10), indicating that they may have experienced more active interactions than the molecular streamers south and east
The exact form of the interactions is not yet clear.

The inflow of molecular gas from exterior gas streamers may explain the mixture of the colder and denser gas with the warmer and more diffuse gas in the western part of the CND (Montero-Casta{\~n}o et al. 2009). 
The Northeast Lobe, with an incomplete--ring shape structure, shows a high velocity gradient and a high velocity dispersion, resembling a localized swirling motion (Figure \ref{fig_mnt12}). 
The overall impression is that the CND is the confluence of many incoming filaments. 
The filaments themselves appear spiral--arm like, winding counterclockwise around the Galactic center.

\section{Discussion}
\label{chap_summary}
Our observations suggest that inflow from $>$20 $pc$ scale in radius, could be depositing molecular mass into the central few parsec. 
The dominant motions of molecular gas approximately follow Keplerian orbits. 
At the scale we observed, the orbits of the exterior gas streamers may vary in size, inclination, and eccentricity.
Some of those orbits are well out of the Galactic plane, such that we see very different orbital shapes upon projection.
The resolved overall rotational motions, the asymmetrical morphology, and the incomplete arm structures with prominent gaps, are consistent with tidal effects demonstrated in numerical simulations (as can be illustrated by a scaled--up version of Figure \ref{fig_mnt0}C in Bonnell \& Rice 2008).
With the assumption of a 5$\times$10$^{6}$\,$M_{\odot}$ point mass located inside the CND, the critical density for the molecular gas structure to be tidally stable is:
\begin{equation}
n_{H_{2}}(|\textbf{x}|/pc)=3.59\times10^{7}\left[ \left(\frac{|\textbf{x}|}{pc}\right)^{-3} \right] (\mbox{cm}^{-3}),
\end{equation}
yielding $n_{H_{2}}(2.5)=2.3\times10^{6}$\,cm$^{-3}$, $n_{H_{2}}(5.0)=0.29\times10^{6}$\,cm$^{-3}$, $n_{H_{2}}(7.5)=0.085\times10^{6}$\,cm$^{-3}$.
At the 4--5 pc radii (e.g. the Southern Arc; Figure \ref{fig_mnt0}, 3), structures traced by CS 1--0 and $^{12}$CO 3--2 but not by the higher excitation transitions, are marginally unstable against the tidal effect. This critical density becomes lower than $n^{\mbox{\scriptsize{CS}}}_{c}$ beyond a $\sim$9\,pc radius.

Inflowing molecular gas following eccentric Keplerian orbits will move outward after passing perihelion. 
With our current data, we cannot distinguish whether the motions in the filaments are inward or outward, as we measure only the radial motions along the line--of--sight. 
The impact of the SgrA East supernova remnant (Serabyn et al. 1992; Lee et al. 2003; Tsuboi et al. 2009), may account for the blue--shifted motions both to the north and to the south, as well as the enhanced linewidths in the northeast and northwest corners of the map (Figure \ref{fig_mnt12}).	
We note the similar issue of whether the expansional motions should be attributed to the explosive expulsion or the eccentric orbital motion in interpreting the Galactic 3 $kpc$ arms (Dame \& Thaddeus 2008; see also Sawada et al. 2004).

Hot gas has been detected inside of the CND, so that the sharp inner edge of the CND may not be due to dynamics as might be expected for a stabilized ring/disk geometry. 
While the ionized mini--spiral arms may be a continuation of outer molecular arms, the bulk of the mass still resides outside of the central parsec region, perhaps held up by the specific angular momentum as well as the luminosity and winds from the central young star cluster. 
A schematic model for the proposed scenario is shown in Figure \ref{fig_schematic}. 
A compatible schematic model for the kinematic processes in the inner $<$2 $pc$ region is presented in Yusef-Zadeh et al. (2000).

\acknowledgments
The GBT and SMA data are from projects GBT11BÐ050 and SMA2011AÐS085, which are parts of the integrated state--of--art imaging project KISS: \underline{K}inematic Processes of the Extremely Turbulent \underline{IS}M around the \underline{S}upermassive Black Holes. 
We acknowledge financial support from ASIAA and NRAO. 
We thank Toney Minter, Glen Langston, and David T. Frayer for assisting the GBT observations. 
We thank Glen Petitpas and Nimesh Patel for supporting the SMA observations.  
We acknowledge Ray Blundell, Sergio Mart\'{i}n, and Jessica Lu for useful comments or suggestions.
{\it Facilities:} \facility{SMA, GBT}

\end{document}